# Spin Dynamics of $La_2CuO_4$ and the Two-Dimensional Heisenberg Model

Anders W. Sandvik[(a),(b)] and Douglas J. Scalapino
*Department of Physics, University of California, Santa Barbara, CA 93106*
(October 28, 1994)

The spin-lattice relaxation rate $1/T_1$ and the spin echo decay rate $1/T_{2G}$ for the 2D Heisenberg model are calculated using quantum Monte Carlo and maximum entropy analytic continuation. The results are compared to recent experiments on $La_2CuO_4$, as well as predictions based on the non-linear $\sigma$-model.

The $CuO_2$ planes of the undoped high-$T_c$ cuprates are good physical realizations of the two-dimensional (2D) antiferromagnetic Heisenberg model. [1] The mapping of this lattice model onto the non-linear $\sigma$-model (nl$\sigma$m) in 2+1-D has led to detailed predictions for various experimentally measurable quantities. [2–4] For $T < 600K$ (above the 3D ordering temperature) the correlation length of $La_2CuO_4$ grows exponentially as the temperature is lowered. [5] The behavior is in close agreement with Quantum Monte Carlo results for the 2D Heisenberg model with a nearest-neighbor coupling $J \approx 1500K$, [6] and corresponds to the the nl$\sigma$m in the low-temperature "renormalized classical" (RC) regime. [2] It was recently suggested [3,4,7,8] that the high-temperature behavior of the cuprates corresponds to the "quantum critical" (QC) regime of the nl$\sigma$m, where the leading temperature dependence of the inverse correlation length is linear. Experimental evidence supporting this scenario has been provided by Imai *et al.*, who measured the spin-lattice relaxation rate $1/T_1$ [9] and the gaussian component of the spin-echo decay rate $1/T_{2G}$ [10] at temperatures as high as $T = 900K$. In particular, Imai *et al.* found that $1/T_1$ and the ratio $T_1 T/T_{2G}$ were both temperature independent at high temperatures, as predicted for the QC regime. [8] These experiments were recently repeated by Matsumura *et al.*. [11] Their results for $1/T_1$ are almost identical to the earlier ones, but for $1/T_{2G}$ the temperature dependence obtained is different at high temperature, causing $TT_1/T_{2G}$ to be temperature dependent, in disagreement with the QC scenario. In addition to this discrepancy, an open question is the reason for the absence of the minimum in $1/T_1$ at $T \approx 750K$, theoretically predicted by Chakravarty and Orbach. [12] In order to settle these questions we have calculated both $1/T_1$ and $1/T_{2G}$ for the 2D Heisenberg model using quantum Monte Carlo simulation and the maximum entropy analytic continuation method. This enables us to compare directly the spin dynamics of the Heisenberg model and $La_2CuO_4$, as well as to assess rigorously the accuracy of the predictions based on the nl$\sigma$m. Lower temperatures can be reached than with high-temperature series expansions [15] and calculations on small clusters, [16] and the approximations necessary with these methods can be avoided.

We find that $1/T_{2G}$ for the 2D Heisenberg model decreases faster than the rate measured for $La_2CuO_4$ by Imai *et al* above $750K$. The temperature dependence is $\sim T^{-2}$ for $0.45 < T/J < 1$, in disagreement with the QC prediction $\sim T^{-1}$. Taking into account the temperature dependence of the spin wave velocity [17] slightly improves on the agreement. For $1/T_1$ our results are in good agreement with the experiments. We also find that while $1/T_1$ exhibits a minimum for a local contact hyperfine coupling of the type used by Chakravarty and Orbach [12], this minimum is absent when the experimentally known on-site and near-neighbor interaction is used.

We study the standard 2D Heisenberg hamiltonian

$$\hat{H} = J \sum_{i=1}^{N} \sum_{\delta} \vec{S}_i \cdot \vec{S}_{i+\delta}, \qquad (1)$$

where $\vec{S}_i$ is a spin-$\frac{1}{2}$ operator, and $\delta$ runs over the nearest neighbors of site $i$. The spin-lattice relaxation rate and the spin-echo decay rate for a given nucleus provide information on the spin susceptibility through the direct and transferred hyperfine couplings of the nuclear spin to surrounding electronic spins. For a $^{63}$Cu nuclear spin $\vec{I}_0$ at site 0, the coupling to the electronic spins $\vec{S}_i$ is given by the hyperfine hamiltonian [18,19]

$$^{63}\hat{H} = A_\perp (I_0^x S_0^x + I_0^y S_0^y) + A_\parallel I_0^z S_0^z + B \sum_\delta \vec{I}_0 \cdot \vec{S}_\delta. \quad (2)$$

The constants $A_\perp, A_\parallel$ and $B$ are known from Knight shift measurements.

With the external field in the direction $\alpha$, the NMR spin-lattice relaxation rate is given by [20]

$$\frac{1}{T_1} = \frac{1}{N} \sum_{\alpha'} \sum_q |A_q^{\alpha'}|^2 S(q, \omega_N), \qquad (3)$$

where $\alpha'$ denotes the two axes perpendicular to $\alpha$, and $A_q^{\alpha'}$ is the fourier transform of the $\alpha'$ component of the hyperfine coupling. The dynamic structure factor $S(q, \omega)$ is related to the imaginary part of the spin susceptibility; $S(q, \omega) = \chi''(q, \omega)/(1 - e^{-\beta \omega})$. Since the resonance frequency $\omega_N$ is small compared to $J$, $1/T_1$ effectively measures $S(q, \omega \to 0)$, averaged with the hyperfine form factor $|A_q^{\alpha'}|^2$. In terms of the inverse fourier transform $S_{mn} = S(m\hat{x} + n\hat{y}, \omega \to 0)$ of $S(q, \omega)$, $1/T_1$ with the external field perpendicular to the $CuO_2$ planes is given by

$$^{63}\left(\frac{1}{T_1}\right)_\perp = 2(A_\perp^2 + 4B^2)S_{00}$$
$$+ 16A_\perp B S_{10} + 16B^2 S_{11} + 8B^2 S_{20}. \qquad (4)$$

This is the rate measured in the experiments by Imai et al. and Matsumura et al.. $S_{mn}(\omega)$ can be obtained from the imaginary-time correlation function

$$C_{mn}(\tau) = \langle S^z_{m\hat{x}+n\hat{y}}(\tau) S^z_0(0)\rangle, \qquad (5)$$

where $S^z_{\vec{r}}(\tau) = e^{\tau\hat{H}} S^z_{\vec{r}} e^{-\tau\hat{H}}$, by inverting the relation

$$C_{mn}(\tau) = \frac{1}{\pi}\int_{-\infty}^{\infty} d\omega S(m\hat{x}+n\hat{y},\omega)e^{-\tau\omega}. \qquad (6)$$

Here $C_{mn}(\tau)$ is computed using quantum Monte Carlo simulation, and the inversion of (6) is carried out with the maximum entropy method. [13,14]

The rate $1/T_{2G}$ is related to the interactions between the nuclear spins. The coupling (2) leads to an indirect nuclear spin-spin interaction, which dominates the direct dipole-dipole interactions. For the external magnetic field applied perpendicular to the $CuO_2$ planes, Pennington and Slichter derived the following expression for $1/T_{2G}$: [21]

$$^{63}\left(\frac{1}{T_{2G}}\right)_\perp = \left(\frac{0.69}{2\hbar}\sum_i J_z^2(\vec{x}_i)\right)^{1/2}. \qquad (7)$$

Here $J_z(\vec{x}_i)$ is the $z$-component of the induced interaction at distance $\vec{x}_i$, given by

$$J_z(\vec{x}_i) = A_\| F_z(\vec{x}_i) + B\sum_\delta F_z(\vec{x}_{i+\delta}), \qquad (8)$$

with

$$F_z(\vec{x}_i) = -\frac{1}{2}\left(A_\| \chi(\vec{x}_i) + B\sum_\delta \chi(\vec{x}_{i+\delta})\right), \qquad (9)$$

where $\chi(\vec{x}_i)$ is the static response at separation $\vec{r} = \vec{x}_i$, given by the Kubo formula

$$\chi(\vec{x}_i) = \int_0^\beta d\tau \langle S^z_i(\tau) S^z_0(0)\rangle. \qquad (10)$$

The factor 0.69 in (7) is the natural abundance of the $^{63}Cu$ isotope.

We have used a recently improved version of the Handscomb quantum Monte Carlo technique [22] to calculate the necessary correlation functions. Unlike standard methods, [23] this technique is free from the systematical errors associated with the Trotter break-up. For the analytic continuation of the imaginary-time data necessary to obtain $1/T_1$, we have implemented the so called "classic" maximum entropy procedure as described in a recent paper by Jarrell and Gubernatis. [14] We have studied systems of $N = 64 \times 64$ spins with periodic boundary conditions, at temperatures $T/J = 0.25 - 1.0$. At these temperatures the correlation length is smaller than the lattice size, and there are virtually no finite size effects.

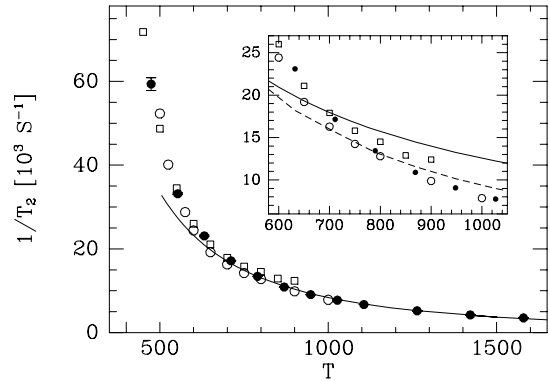

FIG. 1. Monte Carlo results for $1/T_{2G}$ (solid circles) and experimental results by Imai et al. [10] (open squares) and Matsumura et al. [11] (open circles). The solid line in the main figure is of the form $\sim T^{-2}$. Inset: Same as the main figure with the QC prediction by Chubukov et al. [24] (solid curve) and including the temperature dependence of the spin wave velocity (dashed curve).

The calculation of $1/T_{2G}$ is straight-forward, as it involves only the static susceptibility (10). We use the relation $A_\| = -4B$, experimentally known to hold quite accurately. [18,19] We are then left with $J$ and $B$ as fitting parameters, that can be checked against other experiments. The best agreement with the experimental data for $1/T_{2G}$ is obtained with $J = 1580K$ and $B = 3.4 \times 10^{-7}$ eV ($\approx 37$ kOe/$\mu_B$), both consistent with other estimates. [5,6,19] The Monte Carlo results with these parameters are shown in Figure 1, along with the experimental data. Although the over-all agement is good, a notable feature is that for $T > 750K$ the data of Imai et al. [10] is flatter than both the Monte Carlo results and those of Matsumura et al.. [11] This flatness cannot be reproduced for the Heisenberg model with any reasonable values of $J$ and $B$ and, if correct, must be associated with physics not described by this model alone. On the other hand, the data of Matsumura et al. is well reproduced at high temperatures, but exhibits slightly more curvature than the Monte Carlo results and the data of Imai et al. at intermediate temperatures. Fig. 1 also shows the theoretical form derived by Chubukov et al. [24] for the QC regime, which for the hyperfine couplings used here becomes

$$\frac{1}{T_{2G}} \approx 0.491\, \xi(T) \times 10^4 \text{ s}^{-1}, \qquad (11)$$

where the QC correlation length is given by [3,4]

$$\xi = c/(1.038\, T) \qquad \text{(QC regime)}, \qquad (12)$$

and the spinwave velocity $c \approx 1.68$. [27] As noted by Chubukov et al., [24] the over-all magnitude of $1/T_{2G}$ is well reproduced with this formula, but the slope is not. Actually, the Monte Carlo results for $1/T_{2G}$ in the regime $0.45 < T < 1$ is very well described by a $1/T^2$ behavior; a quite significant deviation from (11). Elstner et al. [17] recently pointed out that the leading lattice corrections to the nl$\sigma$m can be taken into account via a temperature dependent spinwave velocity. The velocity calculated from Monte Carlo results for the static structure factor and the static susceptibility agrees well [25] with the high-temperature series expansion results by Elstner et al., [17] and when used in Eq. (12) slightly improves the agreement with the Monte Carlo results for $1/T_{2G}$.

We now turn to the calculation of $1/T_1$, which is more complicated as it relies on a numerical analytic continuation of imaginary time correlation functions. For the local correlation function $C_{00}(\tau)$ the relative statistical errors in our data is typically as low as $10^{-4}$, and the continuation of this quantity is relatively stable. For $C_{10}$, $C_{11}$ and $C_{20}$ the relative errors are typically on the order of $10^{-3}$ and an accurate determination of $1/T_1$ using the full extended hyperfine couplings is therefore more difficult than with a strictly local one.

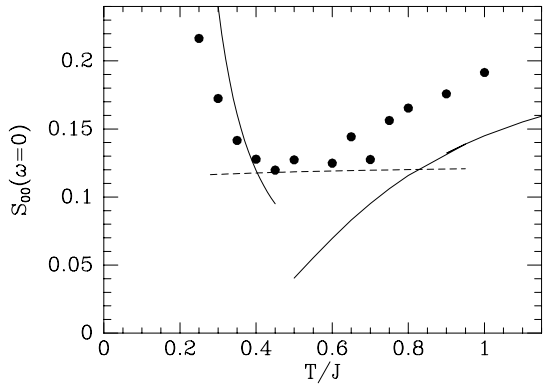

FIG. 2. Maximum entropy results for $S_{00}$ vs T (solid circles) compared to the RC form (13) and the high-temperature form (15) (solid curves). The dashed curve is the QC form (16).

In Figure 2, the $\omega \to 0$ limit of $S_{00}$ is graphed versus the temperature. For a strictly local coupling, this quantity is proportional to $1/T_1$. Repeating the analytic continuation procedure for different subsets of the Monte Carlo data, we estimate the statistical errors to be approximately 5-10% (any bias due to the maximum entropy procedure itself is of course not captured this way). A broad minimum around $T/J = 0.5$ is observed, in good agreement with the prediction by Chakravarty and Orbach, [12] who deduced this feature by contrasting the behavior of $S_{00}(\omega \to 0)$ in the RC regime and the high-temperature limit. The RC expression is [12,4]

$$S_{00}^{RC}(\omega \to 0) = \frac{\lambda N_0^2}{\sqrt{6}} \frac{\xi}{c} \left(\frac{T}{2\pi\rho_s}\right)^{3/2} \left(\frac{1}{1+T/2\pi\rho_s}\right)^2, \quad (13)$$

where the correlation length is given by [2,26]

$$\xi = \frac{e}{8} \frac{c}{2\pi\rho_s} \left(1 - \frac{T}{4\pi\rho_s}\right) e^{2\pi\rho_s/T} \qquad \text{(RC regime)}. \quad (14)$$

The spin-stiffness $\rho_s \approx 0.18$ [27,28] and the ordered moment is $N_0 \approx 0.31$. [28] The constant $\lambda$ has not been calculated rigorously, but an estimate based on fitting the nl$\sigma$m scaling forms to numerical results is $\lambda N_0^2 = 0.61$. [12,4] The rather poor agreement with our result for $S_{00}$ shown in Figure 1 indicates that this value is too large. It should be noted, however, that even the lowest temperatures studied here correspond to the cross-over regime to RC behavior, [17] and perfect agreement with the RC expression cannot be expected. The high-temperature form is [12]

$$S_{00}^{HT}(\omega \to 0) = \frac{(\sqrt{\pi}/8) e^{-(1/2T)^2(1+1/4T)}}{\sqrt{1+1/4T+O[T^{-2}])}} \quad (15)$$

and deviates from the Monte Carlo data by 25% at $T/J = 1$. Clearly terms of order $T^{-2}$ in Eq. (15) might be important for $T \leq 1$. In the regime $0.4 < T/J < 0.7$ $S_{00}(\omega \to 0)$ appears to be rather flat, as predicted for the QC regime. The expression derived by Chubukov et al. is, [4]

$$S_{00}^{QC}(\omega \to 0) = \frac{N_0^2}{\rho_s}\left(\frac{3T}{2\pi\rho_s}\right)^\eta R_1, \quad (16)$$

where the the 3D classical Heisenberg exponent $\eta \approx 0.03$. Deep inside the QC regime $R_1$ is a constant, for which Chubukov et al estimated $R_1 \approx 0.22$ (there are certain complications in estimating $R_1$). [4] Eq. (16) with this value of $R_1$ describes the behavior in the intermediate temperature regime quite well.

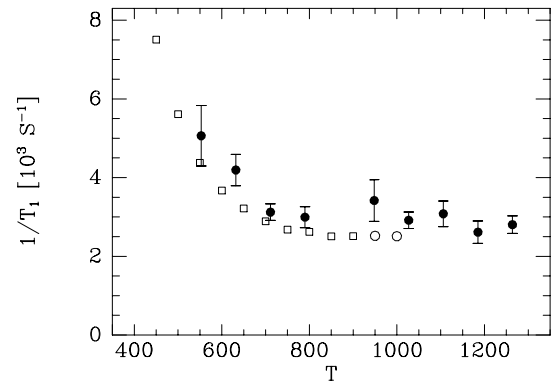

FIG. 3. Maximum entropy results for $1/T_1$ vs T (solid circles) compared to the experimental results by Imai et al. [9] (open squares) and Matsumura et al. [11] (open circles).

The minimum in $1/T_1$ has not been observed experimentally. [9,11] In Figure 3 we show results obtained with the full hyperfine coupling (2), using the same values of $J$ and $B$ as in the fit to $1/T_{2G}$ in Figure 1. For $A_\perp/B$ we take the experimental value 0.84. [18,19] The statistical errors are rather large, as discussed above, but a clear difference from the temperature dependence of Figure 2 can be noted, and the agreement with the results by Imai et al. [9] and Matsumura et al. [11] is reasonably good ( an even better agreement is obtained with a slightly smaller $A_\perp/B$). In particular, $1/T_1$ is temperature independent at high temperatures and the minimum found above for $S_{00}(\omega \to 0)$ at $T \approx 800K$ is not reproduced.

To summarize our results, we note that $1/T_{2G}$ for the Heisenberg model agrees well with the experimental results for $La_2CuO_4$ by Imai et al. [9,10] and Matsumura et al. [11] for $T < 750K$. However, for $T > 750K$ the Heisenberg result decays considerably faster than the data by Imai et al.. The temperature dependence is close to $1/T^2$ in a wide temperature regime. For $1/T_1$ our results are also in reasonable agreement with the experiments and, in particular, is almost temperature-independent at high temperatures, as predicted for the QC regime. [4] As a consequence of the bahavior of $1/T_{2G}$, the ratio $T_1T/T_{2G}$ is not constant at high temperatures, in contradiction to the QC prediction. [8] It appears that the implications of the high-temperature results for $1/T_{2G}$ by Imai et al. have to be reconsidered.

In a recent paper, Elstner et al. [17] found that pure QC behavior for the correlation length of the 2D Heisenberg model can be observed only above $T/J \approx 0.6$, and that lattice effects also become important at these temperatures. In the temperature regime $0.4 < T/J < 0.6$ there is a cross-over to RC behavior. In view of these results the poor agreement with QC scaling for $1/T_{2G}$ found here is perhaps not surprising. On the other hand, $1/T_1$ does exhibit QC behavior for $T > 700K$. It should also be noted that the uniform susceptibility exhibits strikingly accurate QC behavior at *lower* temperatures. [3,4] Hence, it is clear that effects of the proximity to the critical point are manifest in the 2D Heisenberg model. However, the size and location of the regime where QC behavior can be observed depends strongly on the quantity considered, and there does not seem to exist a temperature regime where the QC scaling formulas can be applied universally.

We would like to thank A. Chubukov, H. Monien, R.R.P. Singh, and A. Sokol for very helpful conversations. This work is supported by the Department of Energy under Grant No. DE-FG03-85ER45197.

[a] Present address: National High Magnetic Field Laboratory, 1800 E. Paul Dirac Dr., Tallahassee, FL 32306.
[b] On leave from Deptartment of Physics. Åbo Akademi, Åbo, Finland.

[1] See, e.g., E. Manousakis, Rev. Mod. Phys. **63**, 1 (1991).
[2] S. Chakravarty, B.I. Halperin, and D.R. Nelson, Phys. Rev. Lett. **60**, 1057 (1988); Phys. Rev. B**39**, 2344 (1989).
[3] S. Sachdev and J. Ye, Phys. Rev. Lett. **69**, 2411 (1992); A. Chubukov and S. Sachdev, Phys. Rev. Lett. **71**, 169 (1993).
[4] A. Chubukov, S. Sachdev, and J. Ye, Phys. Rev. B **49**, 11919 (1994).
[5] G. Shirane et al., Phys. Rev. Lett. **59**, 1613 (1987); Y. Endoh et al., Phys. Rev. B**37**, 7443 (1988); B. Keimer et al., Phys. Rev. B**46**, 14034 (1992).
[6] H.-Q. Ding and M.S. Makivić, Phys. Rev. Lett **64**, 1449 (1990); M.S. Makivić and H.-Q. Ding, Phys. Rev. B**43**, 3562 (1991).
[7] A.J. Millis and H. Monien, Phys. Rev. Lett. **70**, 2810 (1993).
[8] A. Sokol and D. Pines, Phys. Rev. Lett. **71**, 2813 (1993).
[9] T. Imai et al. Phys. Rev. Lett. **70**, 1002 (1993).
[10] T. Imai et al. Phys. Rev. Lett. **71**, 1254 (1993).
[11] M. Matsumura, H. Yasuoka, Y. Ueda, H. Yamagata, and Y. Itoh, ISSP Report A2847 (1994).
[12] S. Chakravarty and R. Orbach, Phys. Rev. Lett. **64**, 224 (1990).
[13] J.E. Gubernatis et al., Phys. Rev. B**44**, 6011 (1991).
[14] M. Jarrell and J.E. Gubernatis, preprint (1994).
[15] M.P. Gelfand and R.R.P. Singh, Phys. Rev. B**47**, 14413 (1993).
[16] A. Sokol, E. Gagliano, and S. Bacci, Phys. Rev. B**47**, 14646 (1993).
[17] N. Elstner, R.L. Glenister, R.R.P. Singh, and A. Sokol, preprint (1994).
[18] F. Mila and T.M. Rice, Physica C **157**, 561 (1989).
[19] A.J. Millis, H. Monien, and D. Pines, Phys. Rev. B**42**, 167 (1990).
[20] T. Moriya, Prog. Theor. Phys. **28**, 371 (1962).
[21] C.H. Pennington and C.P. Slichter, Phys. Rev. Lett. **66**, 381 (1991).
[22] A.W. Sandvik and J. Kurkijärvi, Phys. Rev. B **43**, 5950 (1991); A.W. Sandvik, J. Phys. A **25**, 3667 (1992).
[23] J. E. Hirsch et al., Phys. Rev. B **26**, 5033 (1982).
[24] A.V. Chubukov, S. Sachdev, and A. Sokol, Phys. Rev. B **49**, 9052 (1994).
[25] A.W. Sandvik and A. Sokol, unpublished.
[26] P. Hasenfratz and F. Niedermayer, Phys. Lett. B **268**, 231 (1991); Z. Phys. B **92**, 91 (1993).
[27] R.R.P. Singh, Phys. Rev. B **39**, 9760 (1989).
[28] J.D. Reger and A.P. Young, Phys. Rev. B **37**, 5987 (1988); U.J. Wiese and H.P. Ying, Z. Phys. B **93**, 147 (1994).